\documentclass{ws-procs9x6}

\begin{document}

\title{APPLICATION OF CHIRAL NUCLEAR FORCES TO LIGHT NUCLEI}

\author{A. NOGGA$^*$}

\address{Institut f\"ur Kernphysik, Forschungszentrum J\"ulich, 52425 J\"ulich, Germany\\
$^*$E-mail: a.nogga@fz-juelich.de}

\begin{abstract}
In these proceedings, we discuss the current status of nuclear bound state 
predictions based on chiral nuclear interactions. Results of ordinary $s$- and $p$-shell 
nuclei and light hypernuclei are shown. 
\end{abstract}


\bodymatter

\section{Introduction}

One main goal of nuclear physics is the understanding 
of the binding energies of nuclei. In the past, it was the aim to relate 
the binding energies to nuclear forces that describe the two-nucleon 
(NN)  scattering data. This, however, can only be part of a 
complete understanding. Finally, it is necessary to related the binding 
energies to QCD.

The most promising approach to establish this relation is chiral effective 
field theory. It enables us to build into the nuclear interaction  
the symmetries of  QCD and allows 
the determination of unknown parameters  by adjustment 
of lattice calculations even at unphysically large quark masses. 
In this way, a direct connection of nuclear binding energies and QCD 
will be established in the future \cite{SavageCD06}. 

At this time, chiral effective theory is an important guideline to identify 
the most important contributions to nuclear interactions and 
to pin down relations of nuclear interactions to other 
strong interaction processes, e.g.  $\pi$N scattering and $\pi$ production. 
In this context,  it is of utmost importance to pin down the structure
of three-nucleon forces (3NF's). Traditional calculations \cite{Nogga2002,Pieper2001a,Navratil1998a}
clearly show that 3NF's are significant for a quantitative description of 
binding energies. Current models can provide correct binding 
for $s$-shell \cite{Nogga2002}, but fail for $p$-shell nuclei \cite{Pieper2001} 
and scattering observables \cite{Sekiguchi2002,Kistryn2005,Ermisch2005} 
(see also contributions of Johan Messchendorp
and Kimiko Sekiguchi to this conference  \cite{MesschendorpCD06,SekiguchiCD06}). 
Extensions of these models can improve some of these failures 
\cite{Pieper2001} in the regime of light nuclei, 
but again some deviations show up for the more complex systems
\cite{PieperElbatalk}.  A reliable extension, however, is the key to get theoretical 
insight into the structure of, e.g., exotic nuclei. 

Therefore, the application of chiral perturbation theory to the nuclear 
bound state problem is of interest to make  nuclear structure 
calculations more reliable.  We will argue below, that such calculations 
are also important to confirm that we correctly extend 
chiral perturbation theory to the non-perturbative nuclear systems. 

In these proceedings, we discuss the current status of bound state 
calculations for light nuclei and hypernuclei. 
In Section \ref{sec:inter}, we define the chiral interactions 
used for the calculations. We then look in detail at the dependence 
of the results on the cutoff necessary for the regularization of the problem
in Section~\ref{sec:cutoff}. Section~\ref{sec:pshell} is devoted to predictions 
for $p$-shell nuclei. Then, we turn to first results for hypernuclei in 
Section~\ref{sec:hyper} and conclude in Section~\ref{sec:concl}.

\section{Chiral nuclear forces}
\label{sec:inter}

For a complete overview, we refer to the recent reviews on chiral nuclear 
interactions \cite{Epelbaum2006a,Bedaque2002c} (see also Evgeny Epelbaum's contribution 
to this conference \cite{EpelbaumCD06}). 
Here we will only give a summary of the main results 
important for the further discussion. 

A direct application of the power counting of chiral perturbation
theory to nuclear systems is not possible. The existence of nuclear 
bound states excludes any non-perturbative approach. Weinberg 
realized that this non-perturbative behavior is caused by an enhancement 
of diagrams with purely nucleonic intermediate states \cite{WEINBERG1991}. 
He classified such diagrams as ``reducible'' and conjectured that the 
power counting of chiral perturbation theory
can be applied to the ``irreducible'' diagrams. These diagrams then need to be summed 
to all orders using a numerical technique, e.g. solving the Lippmann-Schwinger 
equation. In this way, one obtains in a systematic 
way a nuclear interaction  based on a chiral Lagrangian. The interaction kernel 
is expanded in powers of a typical momentum in nuclei 
or the $\pi$ mass (which is a generic small scale $Q$) over the chiral symmetry 
breaking scale $\Lambda_\chi \approx 1$~GeV. 
The power counting justifies straightforwardly 
the common assumption that NN interactions are much more important 
than 3NF's. Higher order forces are even further suppressed. Quantitative 
results \cite{Ordonez1996,Epelbaum2000,Entem2003,Epelbaum2005} 
confirmed the approach for the NN system. 

The more complex few-nucleon systems, however, promise further 
challenges for this approach. Since 3NF's become quantitatively 
important, the few-nucleon observables are
sensitive to subleading parts of the nuclear interactions. 
Especially, the binding energies are sensitive to details of the interaction,
since they are the result of a rather large cancelation of kinetic energy
and potential energy.

In order $Q^3$ first 3NF's appear \cite{Kolck1994}. 
Three topologies exist.  Chiral symmetry relates 
the strength of the the $2\pi$ exchange 3NF to 
corresponding diagrams of the NN interaction and also to $\pi$N scattering. 
A quantitative confirmation that consistent values of the corresponding 
low energy constants (LEC's) (usually referred to as $c_i$) 
can be found for all of these processes is an important confirmation that chiral symmetry 
is realized in nuclear interactions in the way we assume now. 
At this point fairly consistent values have been extracted from NN and 
$\pi$N data \cite{Meissner2006b,Rentmeester2003,Entem2002b,Buttiker2000,Fettes1998}. For
a more conclusive comparison, the extractions have to be more accurate or 
additional insight from few-nucleon systems is required. 

Except for the $2\pi$ exchange 3NF's, chiral effective theory predicts two more 
leading 3NF structures. Here two a priori 
unknown LEC's appear.  The first one determines the strength of 
the $1\pi$ exchange diagram and can in principle be related to 
$\pi$ production in NN scattering \cite{Hanhart2000} 
or weak processes in few-nucleon systems \cite{Gardestig2006b}. 
In practice, such extractions cannot be used at this time, since 
they were performed in frame works that are not consistent with the one used here.  
The second one enters via the 6N contact vertex. It can only be fixed by matching to 
few-nucleon observables. 

\section{Cutoff dependence of nuclear binding energies}
\label{sec:cutoff}

In any order, the chiral potentials are singular interactions. If the singularities 
are attractive, the Hamiltonian becomes unbounded from below. Therefore,
regularization is required before solving the Lippmann-Schwinger or Schr\"odinger 
equation based on chiral interactions. 
Usually the regularization is performed by means of a momentum cutoff $\Lambda$.
The available realizations of chiral interactions mostly use $\Lambda \approx 
500$-$600$~MeV. Here $\Lambda$ is chosen to be below a typical hadron 
mass, e.g. the $\rho$ mass. A quantitative confirmation for this choice is desirable. 
To this aim, it is instructive to study the cutoff dependence in a much larger 
range of $\Lambda$'s. For a few-body problem this was first done 
in \cite{Nogga2005} for the leading order nuclear interaction only. 
It turned out that in leading order, additional contact interactions beyond 
those required by naive dimensional analysis are required to describe the NN data 
cutoff independently \cite{Valderrama2005b,Nogga2005,Birse2006a}. These results triggered 
an ongoing discussion on the proper power counting for chiral nuclear  
interactions (see e.g. the panel discussion in the few-body working group \cite{KolckCD06,MeissnerCD06,ValderramaCD06}). 
Whereas this issue will be of importance to extend the nuclear interaction 
to momenta close to the $\pi$ production threshold \cite{Nogga2006b}, 
where the currently used cutoffs become similar to typical momenta, 
we will argue below that it does not strongly affect progress for 
nuclear structure calculations. The differences discussed 
in the power counting become quantitatively negligible for high order interactions 
and for cutoffs of the order of $500$~MeV. 

\begin{table}[tb]
\tbl {Expected variations of the results with the cutoff 
for different orders of the interaction. $n$ is the order of the missing contact 
interactions, $\Lambda$ the momentum cutoff used and $\Delta V / V$ 
($\Delta E / E$) are the estimated relative variations of the potential (binding)
energy.}{\hspace{3cm} \begin{tabular}{ccclr}
order & $n$    &   $\Lambda$ [MeV]   &   $\Delta V / V $ & $\Delta E / E $\cr
\hline
$Q^0$  & 2     &     500                       &     7  \%                & 70 \% \cr 
$Q^2$  & 4     &     500                       &     0.5 \%                & 5 \% \cr 
$Q^3$  & 4     &     500                       &     0.5 \%                & 5 \% \cr 
$Q^4$  & 6     &     500                       &     0.03 \%              & 0.3 \% \cr 
\hline
$Q^0$  & 4     &     700                       &     0.1 \%                & 1 \% \cr 
\end{tabular} \hspace{3cm}}
\label{tab:cutexp}

\end{table}

It is useful to discuss first our expectations for the cutoff dependence 
of chiral interactions. To get a rough estimate, one can assume that 
the typical small scale (typical momentum or pion mass) is of the order 
$Q \approx   130$~MeV. One can expect that variations 
of the result are absorbed by contact interactions that are not considered 
at a given order. Assuming natural size, these contact interactions 
should give contributions that scale with $(Q / \Lambda)^n$. Given that 
NN contact interactions contribute only in even orders, one finds 
that $n=2$ in leading order ($Q^0$), $n=4$ in next-leading order (NLO,$Q^2$) 
and next-to-next-to-leading order (N2LO, $Q^3$) and $n=6$ for order $Q^4$ 
(N3LO) interactions. $\Lambda \approx 500$~MeV for typical realizations of chiral 
interactions. Table~\ref{tab:cutexp} shows the expected variation of the potential 
energy for $\Lambda = 500$ and 700~MeV. Assuming 
that these missing contributions can be treated in perturbation theory, one 
finds that this is the actual change expected for the binding energy. This, however,
implies that the relative variation of the binding energy becomes quite large.
In the table, we simple assumed that it is one order of magnitude larger than 
the relative variation of the potential. The table also shows that the 
actual size of the cutoff variation is similar for NLO and N2LO 
(at least if only NN interactions are considered) and that the estimate 
strongly depends on the cutoff, especially for higher order interactions.
Only in order $Q^4$, the variation of the binding energy 
with the cutoff is expected to be very small.
It is now interesting to confront these expectations with actual calculations.

\begin{figure}[tb]
\begin{center}
\psfig{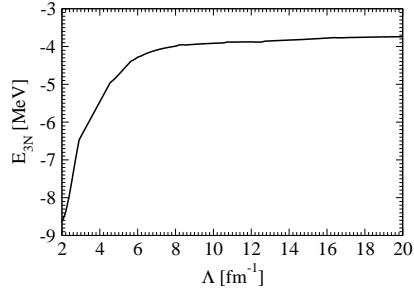}  
\caption{Dependence of the $^3$H binding energy on the cutoff $\Lambda$ for 
the leading order interaction. Thereby additional contact interactions were 
required \cite{Nogga2005}. }
\label{fig:h3cutoff}
\end{center}
\end{figure}

Fig.~\ref{fig:h3cutoff} shows the dependence of the $^3$H binding 
energy on $\Lambda$ for a wide range of cutoffs between 
$\Lambda = 2$-20~fm$^{-1}$
($\approx 400$-5000~MeV). One observes that one can obtain a cutoff independent 
binding energy for large $\Lambda$, once the cutoff dependence of NN predictions 
is removed by promoting counter terms from naively higher orders. 
Contrarily to the effective theory without pions \cite{Braaten2006}, one does not need to 
promote a 3NF to leading order to get cutoff independent results. In the range 
of cutoffs considered, one obtains a variation of the result of approximately
4~MeV consistent with the expected variation of a leading order calculation. 
Note also that in the low cutoff range between 500 and 600~MeV, one finds that 
the binding energy varies quite rapidly by approximately 1.5 MeV. 

\begin{table}[tb]
\tbl{Calculated dependence of the $^3$H binding energy 
              for different chiral interactions. The $Q^0$, $Q^2$, $Q^3$ and $Q^4$ 
              interactions are from Refs.~\cite{Nogga2005,Epelbaum2005,Entem2003}. 
              $\Lambda$ is the momentum cutoff imposed on the Lippmann-Schwinger 
              equation, $\tilde \Lambda$ are cutoffs imposed on internal loops (see 
              Ref.~\cite{Epelbaum2005}). DR notes that loops are dimensionally regulated. 
              }{
\hspace{0.5cm} \begin{tabular}{cc|ccc|c}
order  & $\Lambda$ / $\tilde \Lambda$ [MeV] & $E$ [MeV] & $V$ [MeV] & $\Delta E$ [keV] &   $\Delta E / V$ \cr 
\hline
$Q^0$ & 500 / {\tiny no loops} & -7.50 & -51.8  &   1430  & 3.0 \%  \cr 
            & 600 / {\tiny no loops} & -6.07 &           &             &                      \cr 
\hline             
$Q^2$ & 400 / 700 & -8.46 &                             &   650   & 1.6 \%  \cr 
            & 550 / 700 & -7.81 & -41.1                   &             &                      \cr 
\hline             
$Q^3$ & 450 / 700 & -8.42 &  -38.3                   &   530    & 1.3 \%  \cr 
            & 600 / 700 & -7.89 &                             &             &                      \cr 
\hline             
$Q^4$ & 500 /  DR & -7.84 & -42.3                    &   40     &  0.1 \%  \cr 
            & 600 / DR & -7.80 &                               &           &                      \cr             
\end{tabular} \hspace{0.5cm}}
\label{tab:h3bind}
\end{table}

In view of the fact that so far no higher order realization of chiral 
interactions has been developed that covers a similarly large range of 
cutoffs, it is not possible to study the binding energies in a similar manner 
for the higher orders. An order of magnitude estimate, however, is possible 
by comparing the variation within a small range of 
$\Lambda \approx 500$-$600$~MeV. Neglecting the contribution 
of 3NF's, for which we assume a rather cutoff independent 
contribution to the binding energies, this is shown in Table~\ref{tab:h3bind} 
again for $^3$H for several orders of the chiral expansion. Though the variations 
within the range of cutoffs are somewhat large compared to the estimate 
in Table~\ref{tab:cutexp}, the results still confirm the power counting expectation.  
Quantitatively, the cutoff dependence becomes negligible in order $Q^4$ (N3LO). 

\begin{table}[tb]
\tbl{Strength constants of the $1\pi$ exchange and 6N contact 3NF's (see definition 
       in Ref.~\cite{Epelbaum2002a}). $Q^3$ interactions are  from Ref.~\cite{Epelbaum2005} 
       $Q^4$ interaction is from Ref.~\cite{Entem2003}. 3NF-A and 3NF-B label 
       two sets of parameters that describe the 3N and 4N binding energies equally well.}
       {\hspace{3cm}
\begin{tabular}{l|lrr}
interaction & $\Lambda$ /  $\tilde \Lambda$ [MeV]  &  $c_D$ & $c_E$ \cr 
\hline
$Q^3$ &  450 / 700                &  1.20    & -0.082 \cr
$Q^3$ &  600 / 700                &  -4.27   & -1.25   \cr
$Q^4$-3NF-A &  500 / DR      & -1.11    & -0.66   \cr
$Q^4$-3NF-B &  500 / DR      &  8.14    & -2.02   \cr
 \end{tabular} \hspace{3cm}}
\label{tab:3nflec}
\end{table}

For a quantitative calculation in higher orders, we need to fix the LEC's 
of the 3NF. As discussed above, there are two LEC's unrelated to the 
NN interaction in the leading 3NF's. Therefore, one needs two few-body 
data. Most naturally, we use the $^3$H binding energy in all of your 
determinations of these LEC's. Both, the $^4$He binding energy and the 
doublet neutron-deuteron scattering length are suitable to constrain 
the second LEC. The details of the determinations can be found in 
Refs.~\cite{Epelbaum2002a,Nogga2004}. We have performed the fits for 
combinations of the leading 3NF with the $Q^3$ chiral interactions 
of Ref.~\cite{Epelbaum2005} and the $Q^4$ interaction of 
Ref.~\cite{Entem2003}. The latter combination is not strictly consistent with 
the power counting, since we neglected $Q^4$ contributions to the 3NF 
and the 4NF\cite{Epelbaum2006}. The values of the parameters 
$c_D$ and $c_E$ in the notation of Ref.~\cite{Epelbaum2002a} are given in 
Table~\ref{tab:3nflec} for completeness. 
Note that we find two solutions for the LEC's in conjunction  with
the NN force of Ref.~\cite{Entem2003}, which we are labeled 3NF-A and 3NF-B 
in the following. 

The binding energies of $^3$H and $^4$He are well described by 
the chiral interactions (by construction). 
To confirm that the application of chiral interactions to $s$-shell nuclei 
gives consistent results, it is interesting to 
compare the contribution of 3NF's to the $^3$H and $^4$He 
potential energies to a power counting estimate. The 3NF is formally in order $Q^3$. 
An estimate similar to the one shown in Table~\ref{tab:cutexp} leads to 
an expected contribution of the 3NF of $\approx 2$~\% to the total potential 
energy.  Our calculations show that the 3NF (even the 
various parts of it)  do not contribute more than 7.5 \%  
to the potential energy of $^4$He, which is still in line with the 
power counting estimate. Therefore, we note that the bound 
state calculations for the $s$-shell 
nuclei confirm our expectations from the power counting. 

\section{Predictions for $^6$Li and $^7$Li}
\label{sec:pshell}

With the 3NF's completely  
fixed, we are now in the position to make predictions for $p$-shell 
nuclei. All results for the $p$-shell nuclei 
have been obtained for the $Q^4$ interaction of Ref.~\cite{Entem2003}.
Since the cutoff dependence for an a $Q^4$ chiral interaction 
is negligible small, we will therefore restrict ourselves only 
to one cutoff $\Lambda = 500$~MeV. 

To predict binding energies and spectra of $p$-shell nuclei,
 we need to use a technique for solving the Schr\"odinger 
 equation based on non-local interactions. Here, we will show 
 results based on the ``no-core shell model'' approach (NCSM). 
 Details of the technique and the results 
 for $^7$Li are discussed in \cite{Nogga2006}. Here, it is sufficient 
 to emphasize that excitation energies can be accurately obtained 
 by the NCSM. The accuracy of the binding energy can be estimate to be
 approximately  1~MeV. 

\begin{figure}[tb]
\begin{center}
\psfig{file=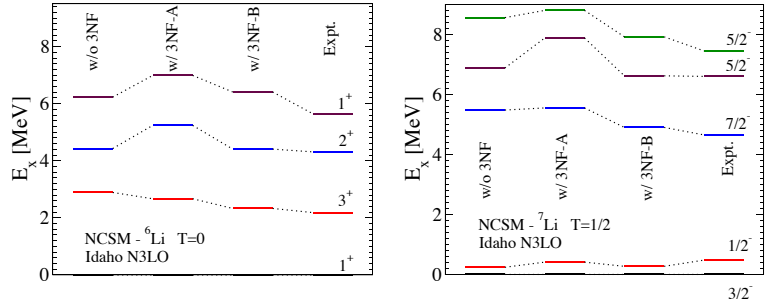,width=4.0in}  
\caption{Spectra of $^6$Li and $^7$Li nuclei based on the chiral interaction 
              of Ref.~\cite{Entem2003} without 3NF or in combination 
              with 3NF-A and 3NF-B. See text for definitions. }
\label{fig:6li7lispectr}
\end{center}
\end{figure}

Since the NCSM results for the spectra are more accurate than the binding 
energies, they are especially important to study the chiral interactions. 
As can be seen in Fig.~\ref{fig:6li7lispectr}, the excitation energies are changed 
by the addition of the 3NF. Note that the two parameter sets 3NF-A and 3NF-B, 
that describe the $s$-shell nuclei equally well, result in different predictions for
$^6$Li and $^7$Li. The expected sensitivity to the structure of the 3NF is confirmed 
by these calculations. For $^6$Li and $^7$Li, we find a consistently better 
description of the spectra for parameter set 3NF-B compared to the 
predictions of 3NF-A and without 3NF. 

\begin{table}[tb]
\tbl{Binding energies $E$ and point proton radii $r_p$ for $^6$Li 
      and $^7$Li. Results for chiral interactions are compared to results 
      based on phenomenological interactions \cite{Pieper2001} and to 
      experimental values that a corrected for the finite size of the protons. 
      }
       {\hspace{2cm}
\begin{tabular}{l|cccc}
interaction  &  \multicolumn{2}{c}{$^6$Li} & \multicolumn{2}{c}{$^6$Li} \cr 
                   & $E$ [MeV]    &  $r_p$ [fm]   & $E$ [MeV]    &  $r_p$ [fm]   \cr
\hline
$Q^4$ -- no 3NF &  -30.0    &    2.20            &  -34.6            &   2.15         \cr 
$Q^4$ -- 3NF-A   &  -32.3    &    2.16            &  -38.0            &   2.11         \cr 
$Q^4$ -- 3NF-B   &  -31.1    &    2.25            &  -36.7            &   2.23         \cr 
\hline 
AV18 -- IL2          &   -32.3     &   2.39            & -38.9             &   2.25         \cr  
AV18 -- Urb-IX     &  -31.1     &    2.57           & -37.5             &   2.33         \cr 
\hline 
Expt.                   &  -32.0      &    2.43            & -39.2            &   2.27         \cr 
 \end{tabular} \hspace{2cm}}
\label{tab:6li7libind}
\end{table}

For the binding energies, the situation is somewhat different. Our results are compiled 
in Table~\ref{tab:6li7libind}. Again 3NF-B improves the description of the radii, but 
both $p$-shell nuclei appear to be underbound. 

This apparent inconsistency of the results deserves some further consideration. 
Here, it is important to note that we have fixed the strength of the 2$\pi$ 
exchange part of the 3NF using the choice of $c_i$ of the NN potential of 
Ref.~\cite{Entem2003}. 
The description of the NN data is not very sensitive to the choice of 
these parameters.  Nuclear matter calculations for low momentum interactions 
including the same 3NF's, however, indicate that a change of the $c_i$'s 
by only 10 \%, 
correcting the $c_D$ and $c_E$ so that the $s$-shell nuclei do not change 
their binding energy, 
may change the binding energy per particle in symmetric nuclear 
matter by 1~MeV \cite{Nogga2004a,Bogner2005}.  It is therefore conceivable 
that a consistent description of binding energies and spectra can be obtained 
by a variation of the $c_i$'s. This needs to be explicitly checked in the future. 

Finally, we note that the addition of 3NF-B, though the relatively 
low cutoffs remove any strong short range repulsion, increases both the binding 
energy and the radii. 

\section{Hypernuclei}
\label{sec:hyper}

Now we turn to hypernuclear binding energies. Recently, Polinder and 
collaborators have developed a first realization of the chiral hyperon-nucleon (YN)
interaction \cite{Polinder2006}. A systematic approach to the problem 
of the YN interaction is badly needed. It will enable us to understand 
the way flavor SU(3) symmetry is broken in nuclear systems. Also the 
impact of hyperons on the nuclear equation of state is possibly 
significant also for astrophysical applications\cite{Lackey2006}, 
but the pour knowledge of the interactions 
of hyperons hinders more insight.
Most of these issues are due to the very scarce set 
of data in the YN sector. Moreover, most of the data are considerably 
above the $\Lambda$N threshold. Therefore, even the scattering lengths 
for $\Lambda$N scattering are essentially unknown (for a discussion on 
the current status see e.g. \cite{Gasparyan2004,Gasparyan2005}).  
  
Current models of the YN interaction \cite{Haidenbauer2005,Rijken1999,Rijken2006} 
do all describe the available YN data, but predictions for non-measured observables 
vary very strongly. They also fail to describe the measured  
binding energies of the light hypernuclei \cite{Nogga2002a} and, therefore,
a more systematic insight into the YN interaction is even more badly needed. 

Also the leading chiral interaction (one Goldstone-boson exchange 
and five non-derivative contact interactions) has been fitted to the scarce data 
base for the YN system. Additionally, the scattering lengths have been 
constrained, so that the $^3_\Lambda$H binding energy is in agreement 
with the experimentally know value of $E=-2.35$~MeV. Thereby, the cutoff 
was varied between 550 and 700~MeV (for details see Ref.~\cite{Polinder2006}).
Contrarily to common expectation, the resulting $\Lambda$N cross section 
at low energies was much smaller than traditional models predict. 
In view of this surprisingly weak interaction, it is astonishing to find 
$^3_\Lambda$H binding energies in agreement with experiment. 
  
\begin{table}[tb]
\tbl{$\Lambda$ separation energies of the $0^+$ ($E_{sep}(0^+)$) 
      and $1^+$ ($E_{sep}(1^+)$) states and their difference $\Delta E_{sep}$ 
      for $^4_\Lambda$H and the 
      difference of the separation energies for the mirror hypernuclei 
      $^4_\Lambda$He and $^4_\Lambda$H (CSB-$0^+$ and CSB-$1^+$). Results for the 
      chiral YN interaction for various cutoffs $\Lambda$ are compared 
      to results for two phenomenological models \cite{Haidenbauer2005,Rijken1999}
      and the experimental values.}
       {\hspace{0cm}
\begin{tabular}{l|cccc|ccc}
$\Lambda$ [MeV] & 500 & 550 & 650 & 700 & J\"ulich 05 & Nijm SC97f & Expt. \cr 
\hline 
 $E_{sep}(0^+)$ [MeV] & 2.63 & 2.46 & 2.36 & 2.38 & 1.87 & 1.60 & 2.04 \cr 
 $E_{sep}(1^+)$ [MeV] & 1.85 & 1.51 & 1.23 & 1.04 & 2.34  & 0.54 & 1.00 \cr 
 $\Delta E_{sep}$ [MeV] & 0.78 & 0.95 & 1.13 & 1.34 & -0.48 & 0.99 & 1.04 \cr 
 \hline 
 CSB-$0^+$ [MeV] & 0.01 & 0.02 & 0.02 & 0.03      & -0.01 & 0.10 & 0.35 \cr 
 CSB-$1^+$ [MeV] & -0.01 & -0.01 & -0.01 & -0.01 & --- &     -0.01    & 0.24 \cr 
 \end{tabular} \hspace{0cm}}
\label{tab:hypnucl}
\end{table}

It is now important to confront predictions based on the leading 
order chiral interaction for the more 
complex hypernuclei $^4_\Lambda$H and $^4_\Lambda$He 
with the data. Two states, the $J^\pi=0^+$ ground and a $J^\pi=1^+$
exited state, of these mirror hypernuclei  are experimentally known. 
Though the YN interactions are not very strongly constrained 
by the YN data, it has proven to be difficult to obtain a consistent 
description of both of the states and also of the well known 
charge dependence of the $\Lambda$ separation energies. It turned out 
that the splittings between the $0^+$ and $1^+$ state and the 
charge dependence of the separation energies are correlated 
with the strong $\Lambda$-$\Sigma$ conversion process.  
Table~\ref{tab:hypnucl} compiles the new results based on
the chiral interaction together with model predictions and the 
experimental values. The dependence on the NN interaction 
is mild \cite{Nogga2002a}. Here, we used the one of 
Ref.~\cite{Entem2003}. The leading order YN interaction 
results in very reasonable $\Lambda$-separation energies. 
The separation energy of the $0^+$  state 
appears to be very cutoff independent, whereas 
the one of the $1^+$ state  is more strongly dependent 
on the cutoff in leading order.
Given that the results are leading order only, they are very encouraging.
NLO calculation will be important to confirm that the agreement 
improves and the cutoff dependence shrinks as we have seen for 
the ordinary nuclei. The leading order calculation 
does not include any charge symmetry breaking contributions 
in the interaction. As a result, the charge dependence of the 
separation energies is very small. (The small non-zero 
contributions is due to the $\Sigma^+$-$\Sigma^-$ mass 
difference and the Coulomb interaction. Both has been included in 
the calculations of Table~\ref{tab:hypnucl}.) The NLO interaction will explicitly include 
first charge symmetry breaking contributions. Therefore, 
an improvement of the leading order results in 
this respect is also conceivable.

\section{Conclusions and Outlook}
\label{sec:concl}

We discussed the results for nuclear binding energies of ordinary 
$s$-shell and $p$-shell nuclei and the lightest hypernuclei based on 
chiral interactions. A special emphasis was on the confirmation 
of the power counting underlying nuclear interactions. 

For the ordinary nuclei, the results of LO, NLO and N2LO 
chiral interactions confirm our power counting expectations. 
We find that N3LO NN interactions give predictions that are 
only insignificantly cutoff dependent. In this order, chiral interaction 
will become quantitatively useful for predictions of nuclear 
binding  energies. In this context, the outcome of an on-going 
discussion on possible promotions of contact interactions 
for formerly higher order will be important, since it might 
enable us to extend the current realizations of chiral interactions 
to slightly larger values of the cutoffs closer to $\Lambda_\chi$
and, thereby, possibly improve the convergence 
of the chiral expansion.

The predictions for $p$-shell nuclei confirm the expected 
sensitivity of the 3NF's. Spectra and binding energies are improved 
by their addition. It will be interesting to allow for 
small variations of the pertinent LEC's $c_i$ to further 
improve the description, especially of the binding energies. 
In this way, an more accurate determination of the $c_i$ 
might become possible. 

We also discussed first results for binding energies of the lightest 
hypernuclei and found that the separation energies of $^3_\Lambda$H 
and  $^4_\Lambda$H can be consistently described. 
Now the extension to NLO will be very interesting. First, it should 
shrink the dependence on the cutoff and enable 
combined fits of the NN and YN interaction. 
Also first charge-symmetry breaking 
terms contribute to the YN interaction in this order. Since 
phenomenological models do not describe the 
charge-symmetry breaking of the $^4_\Lambda$H-$^4_\Lambda$He 
properly, the results of the chiral interaction for this observable 
will be especially interesting.

\section*{Acknowledgments}

I very much thank B. Barrett,  E. Epelbaum, W. Gl\"ockle, 
J. Golak and U. Mei{\ss}ner, P. Navr\'{a}til,  H. Polinder, R. Skibi\'{n}ski, 
R. Timmermans, U. van~Kolck,  J. Vary, H. Wita{\l}a for collaborating 
on the work presented here. The numerical calculations have in part been 
performed on the JUMP and JUBL computers of the NIC, J\"ulich, Germany.

\bibliographystyle{ws-procs9x6}
\bibliography{/Users/andreasnogga/endnote-lib/literatur.bib}

\end{document}